\documentclass[pdflatex,sn-apa]{sn-jnl}

\usepackage{graphicx}%
\usepackage{multirow}%
\usepackage{amsmath,amssymb,amsfonts}%
\usepackage{amsthm}%
\usepackage{mathrsfs}%
\usepackage[title]{appendix}%
\usepackage{textcomp}%
\usepackage{manyfoot}%
\usepackage{booktabs}%
\usepackage{algorithm}%
\usepackage{algorithmicx}%
\usepackage{algpseudocode}%
\usepackage{listings}%
\usepackage{bbm}
\usepackage{natbib}
\usepackage{hyperref}
\usepackage[table]{xcolor}%

\begin{document}

\title[Diff-in-Diff Framework for Policy Drivers]{A Causal Framework for Evaluating Drivers of Policy Effect Heterogeneity Using Difference-in-Differences}


\author*[1]{\fnm{Gary} \sur{Hettinger}}\email{ghetting@pennmedicine.upenn.edu}

\author[2]{\fnm{Youjin} \sur{Lee}}\email{youjin\_lee@brown.edu}

\author[1]{\fnm{Nandita} \sur{Mitra}}\email{nanditam@upenn.edu}

\affil[1]{\orgdiv{Division of Biostatistics}, \orgname{University of Pennsylvania}, \orgaddress{\street{423 Guardian Drive}, \city{Philadelphia}, \postcode{19104}, \state{PA}, \country{U.S.A}}}

\affil[2]{\orgdiv{Department of Biostatistics}, \orgname{Brown University}, \orgaddress{\street{121 South Main Street}, \city{Providence}, \postcode{02912}, \state{RI}, \country{U.S.A.}}}



\abstract{Policymakers and researchers often seek to understand how a policy differentially affects a population and the pathways driving this heterogeneity. For example, when studying an excise tax on sweetened beverages, researchers might assess the roles of cross-border shopping, economic competition, and store-level price changes on beverage sales trends. However, traditional policy evaluation tools, like the difference-in-differences (DiD) approach, primarily target average effects of the observed intervention rather than the underlying drivers of effect heterogeneity. Common approaches to evaluate sources of heterogeneity often lack a causal framework, making it difficult to determine whether observed outcome differences are truly driven by the proposed source of heterogeneity or by other confounding factors. In this paper, we present a framework for evaluating such policy drivers by representing questions of effect heterogeneity under hypothetical interventions and use it to evaluate drivers of the Philadelphia sweetened beverage tax policy effects. Building on recent advancements in estimating causal effect curves under DiD designs, we provide tools to assess policy effect heterogeneity while addressing practical challenges including confounding and neighborhood dynamics.
}




\maketitle

\section{Introduction}\label{sec:intro}

Public policies are crucial in shaping population health and economic outcomes~\citep{PollackPorter2018TheProblems}.
Recently, excise taxes on sweetened beverages have gained popularity as a tool for promoting healthier behaviors and generating government revenue, with implementation in 8 U.S. cities and over 100 countries globally~\citep{Hattersley2023GlobalTaxes}.
While evidence supports reductions in beverage sales and revenue generation in regions implementing a tax, the extent of these effects varies widely both between and within regions, with the factors driving this heterogeneity remaining unclear~\citep{Andreyeva2022OutcomesBeverages}.

For instance, Philadelphia's 1.5 cent per ounce tax on artificially- and sugar-sweetened beverages, implemented in January 2017, led to an estimated 51\% reduction in volume sales~\citep{Roberto2019AssociationSetting}. 
However, an estimated 25-30\% of this decrease was offset by cross-border shopping in neighboring non-taxed counties~\citep{Petimar2022SustainedYears}.
This suggests that differences in store-level factors such as border proximity and nearby economic competition, potentially amplified by tax-related advertising in Philadelphia, may significantly contribute to the observed heterogeneity in tax effects.
Previous studies have linked variations in store-level sales declines to factors such as cross-border shopping accessibility and differential tax pass-through rates~\citep{Cawley2019TheChildren, Hettinger2024DoublyInterference}.
However, understanding the specific causal drivers of policy effect heterogeneity -- rather than unadjusted associations -- is critical for assessing the effectiveness of beverage taxes and other public policies~\citep{Lewis2020AHealth}.

To evaluate these drivers, we can leverage data on how subpopulations were differentially exposed to these drivers.
For example, by examining store proximity to non-taxed regions, we can assess the impact of cross-border shopping accessibility.
Additionally, since the tax was levied on manufacturers instead of consumers, stores had the discretion to adjust their prices, allowing us to explore economic mechanisms such as the impact of store-level pricing decisions and nearby price competition.

When estimating policy effects in observational studies, methodologists commonly use the difference-in-differences (DiD) approach to innately adjust for observed and unobserved confounding factors that differ between intervention and comparison regions and affect outcomes but not outcome \textit{trends} over time~\citep{Angrist2008MostlyEconometrics}. 
Further advancements have refined this approach to adjust for observed confounding differences that affect outcome trends~\citep{Abadie2005SemiparametricEstimators, Santanna2020}.

To evaluate drivers of effect heterogeneity, researchers typically use one of two approaches.
The first is exploratory subgroup analyses, where estimated effect differences between subgroups are interpreted as associations with subgroup characteristics.
While informative, this approach cannot rigorously identify causal mechanisms because it fails to account for confounding differences between subgroups and is prone to bias from subjective subgroup clustering~\citep{Wang2021StatisticalAnalyses}.
The second involves incorporating continuous measures of the driver into models such as a linear two-way fixed effects (TWFE) model.
However, this approach is generally insufficient for identifying causal mechanisms when confounding variables violate the parallel trends assumption and relies on restrictive parametric assumptions that may constrain the possible forms of effect heterogeneity~\citep{Callaway2024Difference-in-differencesTreatment}. 
\cite{Callaway2024Difference-in-differencesTreatment}'s proposed nonparametric estimator for continuous exposures relaxes parametric specifications but still does not distinguish between heterogeneity arising from the driver of interest and other confounding factors.

To rigorously evaluate the drivers of effect heterogeneity, we must determine the effect heterogeneity attributable to a specific driver after adjusting for confounding factors.  
In recent work, \cite{Hettinger2025MultiplyExposures} introduced a robust DiD estimator that nonparametrically estimates causal effect curves for continuous exposures, adjusting for confounders of both intervention and driver.
In this work, we build on these advancements by developing a novel causal framework to assess the \textit{key drivers of effect heterogeneity} motivated by the Philadelphia sweetened beverage excise tax.
Specifically, we frame relevant heterogeneity questions from a fully causal perspective, introduce a new estimand, and address practical challenges such as spatial correlation, economic competition, and joint exposures.
Our framework thus not only enables a deeper analysis of the Philadelphia beverage tax but also serves as a guide for future analyses of effect heterogeneity of other policies.

The paper is organized as follows: Section~\ref{sec:framing} outlines policy-relevant questions on beverage tax effect heterogeneity and conceptualizes these through hypothetical experiments to illustrate the causal framework.
Section~\ref{sec:methods} formalizes these experiments into causal estimands and presents methods to evaluate drivers of interest while addressing practical challenges. 
In Section~\ref{sec:rda}, we apply our framework to investigate the drivers of effect heterogeneity of the Philadelphia beverage tax policy.
Finally, Section~\ref{sec:disc} discusses the broader applicability of this framework and suggests future research directions.

\section{Counterfactual Framing of Policy Drivers under Hypothetical Experiments}\label{sec:framing}

\begin{figure}[htbp]
\centering
    \includegraphics[width=0.78\textwidth]{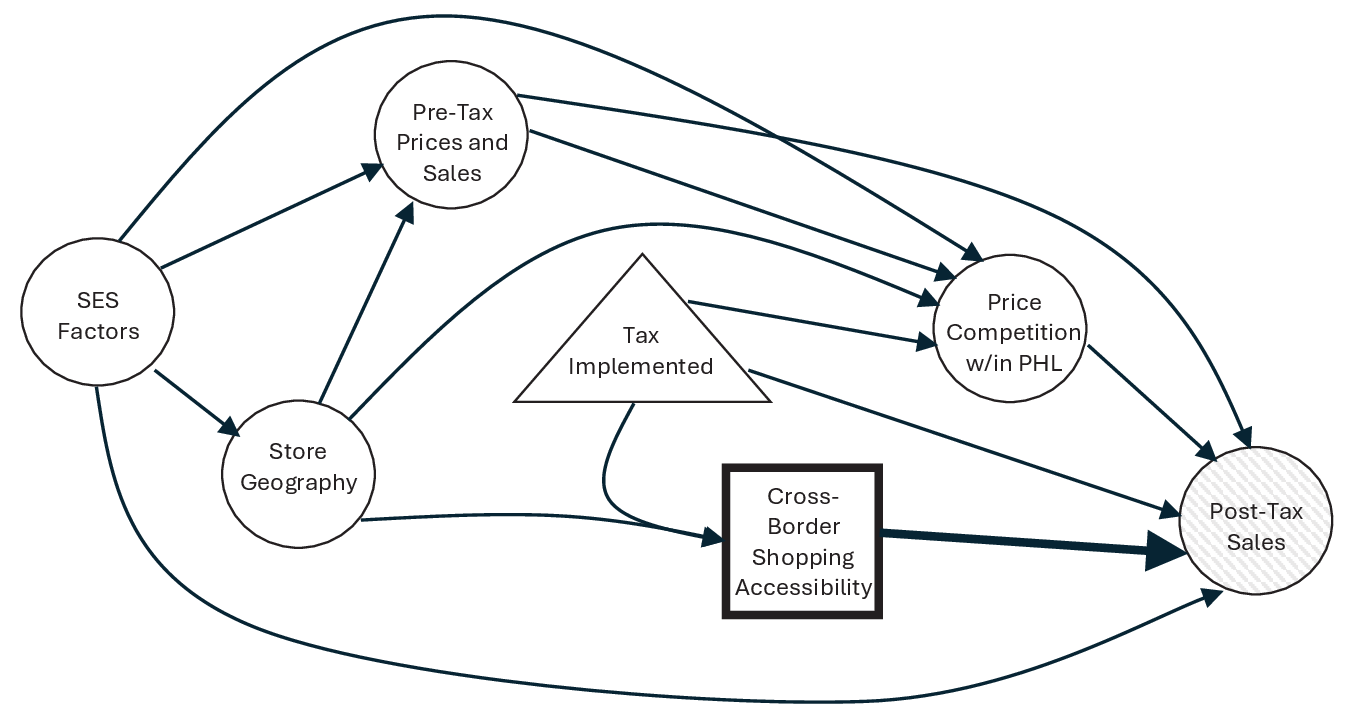}
    \includegraphics[width=0.78\textwidth]{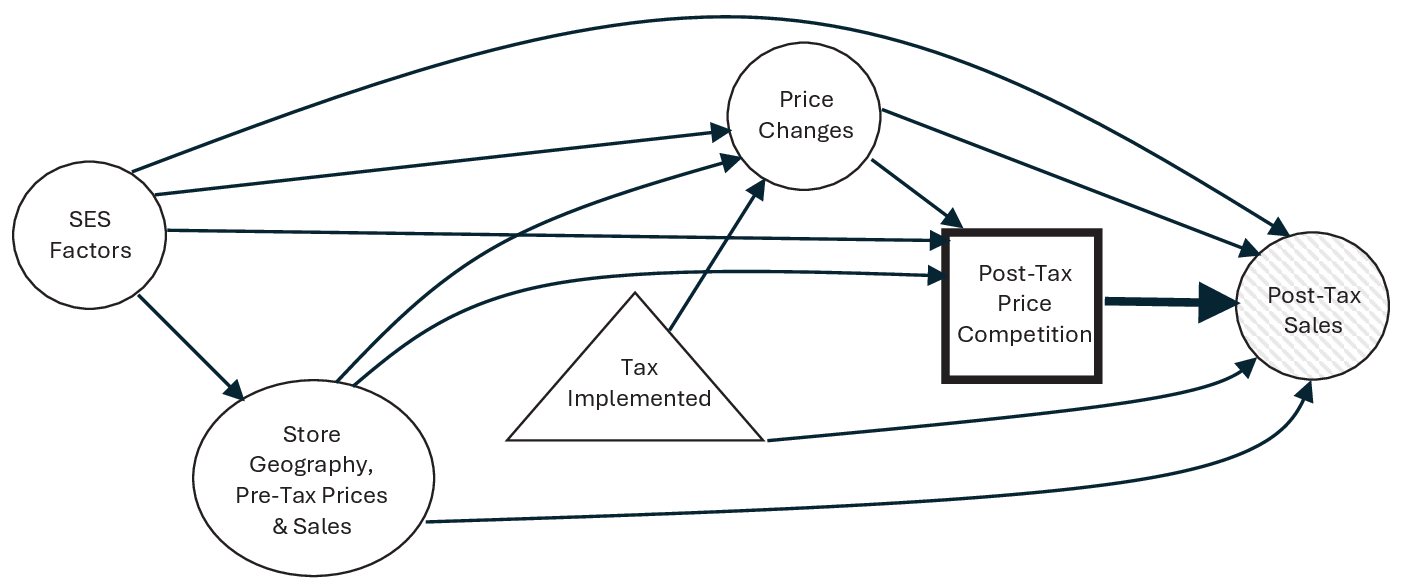}
    \includegraphics[width=0.78\textwidth]{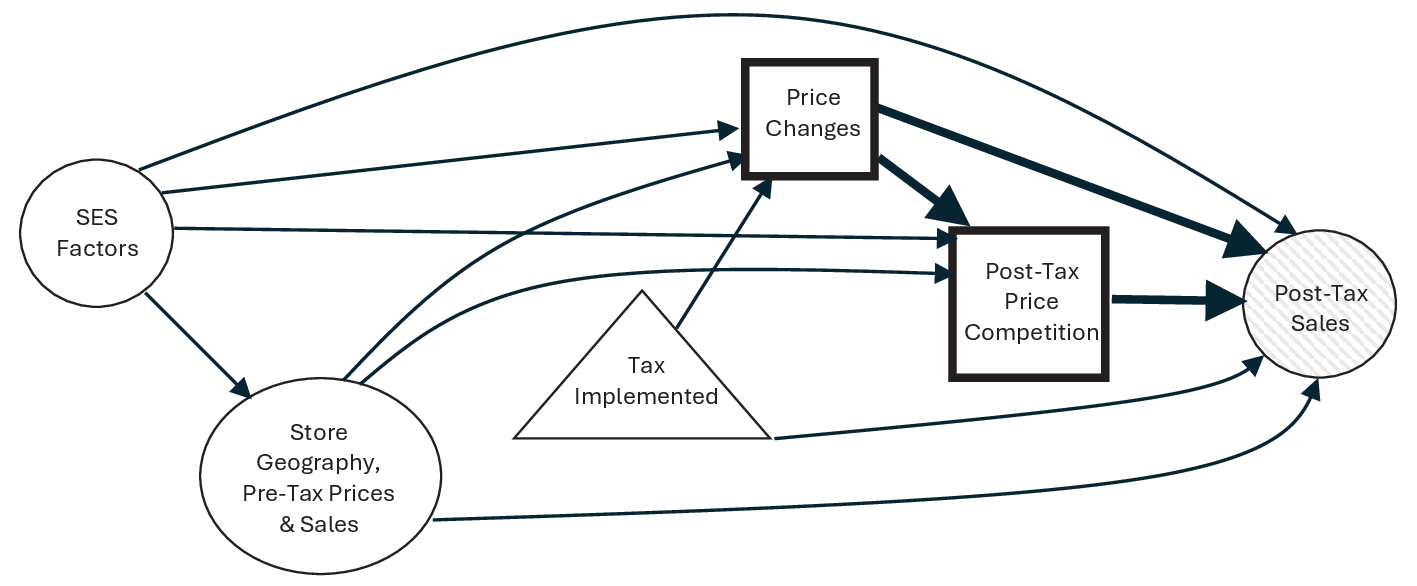}
\caption{Directed acyclic graphs (DAGs) of the mechanisms (square blocks) driving heterogeneous effects on post-tax sales (shaded) at a given Philadelphia store. Confounding factors are represented with unshaded circles. Top: a DAG representing how accessibility to cross-border shopping drives post-tax sales. Middle: a DAG representing how price competition drives post-tax sales. Bottom: a DAG representing how price changes drive post-tax sales. Here, price changes drive sales in part through price competition.}\label{fig:conceptual}
\end{figure}

In this paper, we illustrate our framework by evaluating three key mechanisms through which the tax may differentially affect beverage sales at a Philadelphia store, as depicted in Figure~\ref{fig:conceptual}. 
Specifically, we consider (i) exposure to cross-border shopping, measured by proximity to a non-taxed store, (ii) nearby economic competition, measured by the store's price relative to the neighborhood minimum, and (iii) the store-level pass-through rate, measured by price changes from the year before the tax. 
Notably, these drivers are interrelated and associated with confounders like zip-code level socioeconomic status.

First, we examine the cross-border shopping driver by assessing the role of proximity to a non-taxed region, illustrated by the top Directed Acyclic Graph (DAG) in Figure~\ref{fig:conceptual}.
This mechanism captures how the tax can affect Philadelphia store sales through cross-border shopping to neighboring, non-taxed stores. 
To assess the influence of distance to the border on tax effects, we can imagine a hypothetical experiment where \textit{all} Philadelphia stores were fixed at a distance, $\delta$, from a non-taxed region and evaluate how tax effects vary over different levels of $\delta$.

Second, we explore the role of nearby price competition in driving tax effect heterogeneity, shown by the middle DAG in Figure~\ref{fig:conceptual}. 
Consumers in Philadelphia may be particularly sensitive to post-tax price changes due to the pervasive signage warning of the tax and the city's relatively low socioeconomic status (SES), which may drive them to shop around for better prices and affect sales at individual stores.
To assess the effect of price competition, we can imagine a hypothetical experiment where we fix the minimum beverage price within each store's neighborhood to $\delta$ cents higher than the store's price and analyze how varying levels of price competition alter tax effects.
To operationalize this hypothetical experiment, we must account for associations between price competition and both (1) store-level price changes and (2) border proximity. 

Third, we investigate how store-level pricing decisions influence tax effectiveness,  depicted in the bottom DAG in Figure~\ref{fig:conceptual}. 
In theory, because the tax is placed on the distributor, stores choose how much of the tax to pass through to consumers to optimize profits by using information about their consumer base and past sales, which could influence tax effectiveness.
To assess the impact of this pricing mechanism, we can imagine a hypothetical experiment where stores are randomly assigned a price change, $\delta$, and compare the effect under this hypothetical assignment mechanism to the observed effect when stores independently choose their price changes.
To mimic this experiment, we must also incorporate changes in nearby price competition under this randomized price adjustment.

\section{Methods}\label{sec:methods}

\subsection{Notation}

Consider a policy \textit{intervention} that induces a distribution of \textit{drivers} on the population receiving it.
Our unit of analysis (the store) is the unit at which outcomes are measured and different driver levels could be defined, whereas distinct interventions are often delivered to groups of these units (e.g., regions). 
We index stores by $i=1,...,n$, with each store in either the taxed group ($A_i=1$) or the control group ($A_i=0$). 
The intervention occurs between two time periods, $t\in \{0,1\}$, and the binary time-varying tax status at time $t$ is denoted by $Z_{it}=tA_i$. 
In each period, stores set the price of the product as $\rho_{it}$.

Analogously, units can be differentially exposed to drivers, which may induce tax effect heterogeneity. 
We define exposures to these drivers as follows: distance to the border $B_i \in \mathbbm{B}$, pass-through rate $P_i = \rho_{i1} - \rho_{i0} \in \mathbbm{P}$, and effective price competition $h_{it}(\boldsymbol{\rho_t}) \in \mathbbm{H}$.
Here, $h_{it}(\boldsymbol{\rho_t})$ is a function of the prices of all study units at time $t$, $\boldsymbol{\rho_t}$, similar to the exposure mapping function introduced by \cite{Aronow2013}. 
This function summarizes how stores are exposed to prices at neighboring stores, which we specify as the difference between $\rho_{it}$ and the minimum $\rho_{jt}$ for $j$ in the neighborhood of $i$, where neighborhoods are defined as the directly adjacent zip-codes of $i$. 
We separately define competition within Philadelphia (i.e., not including nearby, non-taxed stores) as $h_{it}(\boldsymbol{\rho_t}|A=1)$.

To distinguish between binary interventions and continuous drivers, we refer to the store's level of exposure to the specific driver as the \textit{dose} it receives.
We observe outcomes such as volume sales, $Y_{it}$, in each period. 
Finally, we denote observed covariates as $\mathbf{X_i}$, which will be used to account for the effects of pre-intervention confounders and other drivers on outcome trends. 

\subsection{Causal Estimands}

In this section, we frame the hypothetical interventions from Section~\ref{sec:framing} as causal parameters, namely contrasts in potential outcomes. 
Depending on the research question, we define potential outcomes, $Y_{it}^{(\boldsymbol{\zeta_i})}$, according to a vector, $\boldsymbol{\zeta_i}$, of interventions and/or drivers subject to hypothetical intervention. 
Expectations in the following formulations are taken across units $i$, and we omit the unit subscript for simplicity.
These estimands will be further demonstrated and interpreted in Section~\ref{sec:rda}. 

\subsubsection*{Average Treatment Effect on the Treated (ATT)}

First, we consider the ATT, a common estimand in policy evaluations: 
\begin{equation}
\label{eqn:att}
ATT := E[Y_1^{(Z_1=1)} - Y_1^{(Z_1=0)}|A=1].
\end{equation}
This estimand asks: ``What would be the average difference in post-tax sales at Philadelphia stores had all stores been taxed compared to not taxed?''
Here, $\boldsymbol{\zeta_i}=Z_{i1}$, since we only manipulate the tax status.

\subsubsection*{Average Dose Effect on the Treated (ADT)}

Second, we consider an effect curve, the ADT, as introduced in \cite{Hettinger2025MultiplyExposures}. 
We define $\mathbf{D_i}$ and $\boldsymbol{\mathcal{D}_{it}}$ analogously to $A_i$ and $Z_{it}$, but in the context of driver doses. 
Specifically, $\mathbf{D_i}$ represents the dose assigned to a store, while $\boldsymbol{\mathcal{D}_{it}} = t\mathbf{D_i}$ denotes the actual driver dose active at time $t$.
Because we do not maniupulate $\boldsymbol{\mathcal{D}_{it}}$ under the control exposure in our hypothetical experiments in Section~\ref{sec:framing}, we define $\boldsymbol{\zeta_i}=(Z_{i1},\boldsymbol{\mathcal{D}_{it}})$ when $Z_{i1}=1$ and $\boldsymbol{\zeta_i}=Z_{i1}$ when $Z_{i1}=0$.
In our study, we will define drivers as uni-dimensional measures like distance to the border ($\mathbf{D_i}=B_i$) and economic competition ($\mathbf{D_i}=h_{i1}(\boldsymbol{\rho_1})$), as well as a two-dimensional joint exposure of price change and economic competition ($\mathbf{D_i}=(P_i, h_{i1}(\boldsymbol{\rho_1}))$). 
Then, the ADT is:
\begin{equation}
\label{eqn:adt}
    ADT(\boldsymbol{\delta}) = E[Y_1^{(Z_1=1,\boldsymbol{\mathcal{D}_{1}}=\boldsymbol{\delta})} - Y_1^{(Z_1=0)} | A=1]
\end{equation}
This estimand asks: ``What would be the average difference in post-tax sales at Philadelphia stores were all stores taxed and exposed to the driver at level $\boldsymbol{\mathcal{D}_1}=\boldsymbol{\delta}$ versus given the control exposure?''

\subsubsection*{Average Treatment Effect of a Dose-Unconfounded Treatment on the Treated (ADUTT)}

Third, we consider the ADUTT:
\begin{equation}
\label{eqn:ADUTT}
ADUTT(f_D) = \int\limits_\mathbbm{D} E[Y_1^{(Z_1=1,\boldsymbol{\mathcal{D}_{1}}=\boldsymbol{\delta})} - Y_1^{(Z_1=0)} | A=1] df_D(\boldsymbol{\delta})
\end{equation}
This stochastic estimand asks: ``What would be the average difference in post-tax sales at Philadelphia stores were all stores taxed but randomly assigned an exposure to $f_D(\boldsymbol{\delta})$ versus given the control exposure?'' 
Without loss of generality, we assume $f_D(\boldsymbol{\delta}) = p_D(\boldsymbol{\delta}|A=1)$ is the distribution of $\mathbf{D}$ among the taxed region, to best align with the observed setting studied by the ATT. 
Thereby, the ADUTT can alternatively be interpreted as the average ADT over the realized distribution of doses.

\subsubsection*{Relative Effect of Dose Assignment (REDA)}

The ADUTT is often most relevant for comparisons to the ATT.
Thus, we define the REDA as:
\begin{equation*}
REDA = (ATT - ADUTT)/ATT
\end{equation*}
The REDA quantifies how the ATT would relatively change if doses were independently assigned under the intervention. 
For example, a REDA of $50\%$ implies that $50\%$ of the ATT is explained by the non-randomness of dose assignment, whereas a REDA of $-50\%$ implies that the intervention effect would be $50\%$ higher under unconfounded dose assignment. 

\subsubsection*{Mathematical Connection Between the ATT and ADUTT}

We can alternatively define the ATT under $\boldsymbol{\zeta_i}=(Z_{i1},\boldsymbol{\mathcal{D}_{i1}})$ as:
\begin{equation*}
    ATT=\int E[Y_1^{(Z_1=1,\boldsymbol{\mathcal{D}_{1}}=\boldsymbol{\delta})} - Y_1^{(Z_1=0)}|A=1,\mathbf{x}] dp_{X,D}(\mathbf{x}, \boldsymbol{\delta}|A=1)
\end{equation*}
This formulation of the ATT, which can be identified and estimated as the previous formulation, integrates over the joint distribution of confounders and doses, $p_{X,D}$, whereas the ADUTT integrates sequentially over the marginal distributions of $X$ and $D$, i.e., $p_X$ and $p_D$:
\begin{equation*}
    ADUTT=\int\limits_\mathbbm{D} \int\limits_\mathbbm{X} E[Y_1^{(Z_1=1,\boldsymbol{\mathcal{D}_{1}}=\boldsymbol{\delta})} - Y_1^{(Z_1=0)} | A=1, \mathbf{x}] dp_X(\boldsymbol{x}) dp_D(\boldsymbol{\delta})
\end{equation*}
This distinction underscores the interpretation of the ADUTT as an effect under \textit{random}, rather than \textit{confounded}, dose assignment.

\subsection{Identification Assumptions}

To identify these causal estimands, we require several (generally untestable) assumptions to map observable data to relevant counterfactuals. 

\subsubsection*{Arrow of Time (No Anticipation)}

We assume potential sales at time $t$ are not influenced by future intervention status or driver dose.
This condition would be violated, for instance, if consumers began altering their shopping habits before the tax was implemented. 
Previous studies have found limited evidence of this behavior in Philadelphia beverage tax data~\citep{Roberto2019AssociationSetting, Hettinger2024DoublyInterference}.

\subsubsection*{Modifed Stable Unit Treatment Value Assumption (SUTVA)}

We require a modified form of SUTVA, where potential outcomes depend on the population-level intervention status, $\mathbf{Z_t}$, and driver levels, $\boldsymbol{\overline{\mathcal{D}}}$ only through the individual unit's intervention and driver status: $Y_{it}^{(\mathbf{z_t}, \boldsymbol{\overline{d}_t})} = Y_{it}^{(z_{it}, \boldsymbol{d_{it}})}$.
This assumption, often trivial in clinical trials, requires careful consideration in policy evaluations. 
For example, when evaluating border proximity as an exposure, we assume that sales at store $i$ in Philadelphia are unaffected by the border proximity of other Philadelphia stores.
However, this assumption becomes less plausible when evaluating price changes, as sales at store $i$ may depend on price changes of nearby stores. Hence, we denote our driver as bi-dimensional, $\mathbf{D_{i}}=(P_i,h_{i1}(\boldsymbol{\rho_1}))$, for this hypothetical intervention, thus assuming that consumers primarily respond to the specified form of price competition ($h_{it}$). 

\subsubsection*{Consistency Assumption}

We assume that potential outcomes are equal to the observed outcomes equal the potential outcomes, $Y_{it}^{(z_{it}, \boldsymbol{d_{it})}} = Y_{it}$, when $Z_{it}=z_{it}$ and $\boldsymbol{\mathcal{D}_{i1}}=\boldsymbol{d_{it}}$. 

\subsubsection*{Positivity Assumption}

This assumption requires all units to have a non-zero probability of being assigned to each relevant intervention status and driver level.
Essentially, it mandates sufficient overlap in covariates across the supports of intervention statuses and driver levels to balance confounders and extrapolate causal effects to the entire treated population.

\subsubsection*{Parallel Trends Assumptions}

Finally, we require two forms of parallel trends:
\begin{enumerate}
    \item \textbf{For the ATT, ADT, ADUTT, and REDA:} A conditional counterfactual parallel trends assumption between treated and control units $E[Y_{1}^{(0)}-Y_{0}^{(0)} | A=1,\mathbf{X}] = E[Y_{1}^{(0)}-Y_{0}^{(0)} | A=0,\mathbf{X}]$. This assumes that non-taxed stores are valid proxies for what would have happened to similar (by $\mathbf{X}$) taxed stores had no tax been implemented.
    \item \textbf{For the ADT, ADUTT, and REDA:} A conditional counterfactual parallel trends assumption among treated units between dose levels~\citep{Hettinger2025MultiplyExposures}: $E[Y_{1}^{(1,\boldsymbol{\delta})}-Y_{0}^{(0)} | A=1,\mathbf{D_{i}}=\boldsymbol{\delta},\mathbf{X}] = E[Y_{1}^{(1,\boldsymbol{\delta})}-Y_{0}^{(0)} | A=1,\mathbf{X}] \text{ for all } \boldsymbol{\delta} \in \mathbbm{D}$. This assumes that taxed stores exposed to driver level $\boldsymbol{\delta}$ are valid proxies for what would have happened to similar (by $\mathbf{X}$) taxed stores had all Philadelphia stores received driver level $\boldsymbol{\delta}$.
\end{enumerate}

When these assumption do not hold, the estimated effects should be interpreted as associations between the intervention/driver and outcome that remain unexplained by observable factors.

\subsection{Estimation Approach} 
\label{sec:methods:estimation}

In this section, we summarize multiply-robust estimators for the $ATT$~\citep{Santanna2020} and $ADT$~\citep{Hettinger2025MultiplyExposures} and introduce
a new estimator for the $ADUTT$, whose efficient influence function has been previously derived~\citep{Hettinger2025MultiplyExposures}.

\subsubsection*{Additional Notation and Key Functions} 

We define the following conditional expectations:
\begin{itemize}
    \item Outcome trends among taxed stores given confounders and driver level:

    $\mu_{1,\Delta}(\mathbf{X}, \mathbf{D}) = E[Y_1 - Y_0 | A=1, \mathbf{X}, \mathbf{D}]$

    \item Outcome trends among the non-taxed stores given confounders:

    $\mu_{0,\Delta}(\mathbf{X}) = E[Y_1 - Y_0 | A=0, \mathbf{X}]$
\end{itemize}

Additionally, we define the following probability functions: 
\begin{itemize}
    \item Probability of being in a taxed region given confounders: $\pi_A(\mathbf{X}) = P(A=1|\mathbf{X})$

    \item Driver level density given confounders: $\pi_D(\mathbf{X}, \mathbf{D}) = p(\mathbf{D} | A=1, \mathbf{X})$

    \item Marginalized driver density: $p(\mathbf{D}|A=1) = \int\limits_\mathbbm{X} \pi_D(\mathbf{x},\mathbf{D}) dp(\mathbf{x}|A=1)$
\end{itemize}

These functions contribute to two composite functions based on the efficient influence functions for the $ATT$ and $ADUTT$ (Web Appendix A):
\begin{align*}
    \xi(\mathbf{X}, A, \mathbf{D}, Y_0, Y_1; \mu_{1,\Delta}, \pi_D) &= \frac{(Y_1-Y_0) - \mu_{1,\Delta}(\mathbf{X}, \mathbf{D})}{\pi_D(\mathbf{X},\mathbf{D})} P(\mathbf{D}|A=1) + \\  & \hspace{10mm} \int\limits_\mathbbm{X} \mu_{1,\Delta}(\mathbf{x}, \mathbf{D}) dP(\mathbf{x}|A=1)  \\
    \tau(\mathbf{X}, A, Y_0, Y_1; \mu_{0,\Delta}, \pi_A) &=  \frac{(1-A)\pi_A(\mathbf{X})[(Y_1 - Y_0) - \mu_{0,\Delta}(\mathbf{X})]}{P(A=1)(1-\pi_A(\mathbf{X}))} + \\
     & \hspace{10mm} \frac{A}{P(A=1)}\mu_{0,\Delta}(\mathbf{X})
\end{align*}

\subsubsection*{Estimation Procedure}

\begin{enumerate}
    \item \textbf{Fit models} for nuisance functions: $\hat{\mu}_{1,\Delta}$, $\hat{\mu}_{0,\Delta}$, $\hat{\pi}_A$,  and $\hat{\pi}_D$. 
    Section~\ref{sec:methods:confound_comp} provides guidance on confounder/exposure definitions, while Section~ \ref{sec:rda:implementation} discusses implementation.
    \item \textbf{Compute unit-specific contributions}, $\hat{\xi}_i$ and  $\hat{\tau}_i$, by plugging empirical data into $\xi$ and $\tau$ for all units. Integrals are approximated via sample means over treated units' covariates for each driver level, $\mathbf{D}$.
    \item \textbf{For the $ADT(\boldsymbol{\delta})$}, fit a non-parametric regression model (we recommend local linear kernel regression) as $\widehat{ADT}(\boldsymbol{\delta}) = \hat{\theta}(\mathbf{D})$, where  $\hat{\theta}(\mathbf{D})$ is obtained by fitting $\hat{\xi}$ as a function of $\mathbf{D}$.
    \item \textbf{Compute final estimates:}
    \begin{enumerate}[a.]
        \item $\widehat{ATT} = \frac{1}{n}\sum\limits_{i=1}^n [ \frac{A}{P(A=1)} (Y_{i1}-Y_{i0}) - \hat{\tau}_i ]$
        \item $\widehat{ADT}(\boldsymbol{\delta}) = \hat{\theta}(\boldsymbol{\delta}) - \frac{1}{n}\sum\limits_{i=1}^n\hat{\tau}_i$
        \item $\widehat{ADUTT} = \frac{1}{n}\sum\limits_{i=1}^n[\hat{\xi}_i - \hat{\tau_i} ]$
        \item $\widehat{REDA} = \frac{\widehat{ATT}-\widehat{ADUTT}}{\widehat{ATT}}$
    \end{enumerate}
\end{enumerate}

Table~\ref{tab:robustness} sumarizes robustness properties. Because $\widehat{ATT}$ depends on $\tau$ but not $\xi$, it only requires $\hat{\mu}_{0,\Delta}$ and $\hat{\pi}_A$ and is consistent if either are correctly specified. 
$\widehat{ADUTT}$, $\widehat{ADT}(D)$, and $\widehat{REDA}$ depend on both $\tau$ and $\xi$, thereby requiring both (i) either $\hat{\mu}_{0,\Delta}$ and $\hat{\pi}_A$ are correctly specified and (ii) $\hat{\mu}_{1,\Delta}$ or $\hat{\pi}_D$ are correctly specified.

\begin{table}[htbp]
\caption{A summary of robustness properties of different estimators under different nuisance function specifications.}\label{tab:robustness}%
\begin{tabular}{@{}cc|cc||cc@{}}
\toprule
\multicolumn{2}{c}{$D$-Functions} & \multicolumn{2}{c}{$A$-Functions} & \multicolumn{2}{c}{Estimators} \\
\cmidrule{1-2}  \cmidrule{3-4} \cmidrule{5-6} \\
$\mu_{1,\Delta}$ & $\pi_D$ & $\mu_{0,\Delta}$ & $\pi_A$ &  $\widehat{ATT}(\delta)$ & $\widehat{ADT}(\delta)$, $\widehat{ADUTT}$, $\widehat{REDA}$ \\
\midrule
\cellcolor{green!40} Good & \cellcolor{green!40} Good & \cellcolor{green!40} Good & \cellcolor{green!40} Good & \cellcolor{green!40} Unbiased & \cellcolor{green!40} Unbiased \\
\cellcolor{green!40} Good & \cellcolor{green!40} Good & \cellcolor{green!40} Good & \cellcolor{red!40} Bad & \cellcolor{green!40} Unbiased & \cellcolor{green!40} Unbiased \\
\cellcolor{green!40} Good & \cellcolor{green!40} Good & \cellcolor{red!40} Bad & \cellcolor{green!40} Good & \cellcolor{green!40} Unbiased & \cellcolor{green!40} Unbiased \\
\cellcolor{green!40} Good & \cellcolor{red!40} Bad & \cellcolor{green!40} Good & \cellcolor{green!40} Good & \cellcolor{green!40} Unbiased & \cellcolor{green!40} Unbiased \\
\cellcolor{red!40} Bad & \cellcolor{green!40} Good & \cellcolor{green!40} Good & \cellcolor{green!40} Good & \cellcolor{green!40} Unbiased & \cellcolor{green!40} Unbiased \\
\cellcolor{green!40} Good & \cellcolor{red!40} Bad & \cellcolor{green!40} Good & \cellcolor{red!40} Bad & \cellcolor{green!40} Unbiased & \cellcolor{green!40} Unbiased \\
\cellcolor{red!40} Bad & \cellcolor{green!40} Good & \cellcolor{red!40} Bad & \cellcolor{green!40} Good & \cellcolor{green!40} Unbiased & \cellcolor{green!40} Unbiased \\
\cellcolor{green!40} Good & \cellcolor{red!40} Bad & \cellcolor{red!40} Bad & \cellcolor{green!40} Good & \cellcolor{green!40} Unbiased & \cellcolor{green!40} Unbiased \\
\cellcolor{red!40} Bad & \cellcolor{green!40} Good & \cellcolor{green!40} Good & \cellcolor{red!40} Bad & \cellcolor{green!40} Unbiased & \cellcolor{green!40} Unbiased \\
\cellcolor{red!40} Bad & \cellcolor{red!40} Bad & \cellcolor{green!40} Good & \cellcolor{green!40} Good & \cellcolor{green!40} Unbiased & \cellcolor{red!40} Biased \\
\cellcolor{red!40} Bad & \cellcolor{red!40} Bad & \cellcolor{green!40} Good & \cellcolor{red!40} Bad & \cellcolor{green!40} Unbiased & \cellcolor{red!40} Biased\\
\cellcolor{red!40} Bad & \cellcolor{red!40} Bad & \cellcolor{red!40} Bad & \cellcolor{green!40} Good & \cellcolor{green!40} Unbiased & \cellcolor{red!40} Biased\\
\cellcolor{green!40} Good & \cellcolor{green!40} Good & \cellcolor{red!40} Bad & \cellcolor{red!40} Bad & \cellcolor{red!40} Biased & \cellcolor{red!40} Biased\\
\cellcolor{green!40} Good & \cellcolor{red!40} Bad & \cellcolor{red!40} Bad & \cellcolor{red!40} Bad & \cellcolor{red!40} Biased & \cellcolor{red!40} Biased \\
\cellcolor{red!40} Bad & \cellcolor{green!40} Good & \cellcolor{red!40} Bad & \cellcolor{red!40} Bad & \cellcolor{red!40} Biased & \cellcolor{red!40} Biased\\
\cellcolor{red!40} Bad & \cellcolor{red!40} Bad & \cellcolor{red!40} Bad & \cellcolor{red!40} Bad & \cellcolor{red!40} Biased & \cellcolor{red!40} Biased\\
\bottomrule
\end{tabular}
\end{table}

\subsubsection*{Inference}

To conduct inference on these parameters, we use a weighted block bootstrapping approach to account for spatial correlation~\citep{EfronTibshirani1993, Lahiri1999TheoreticalMethods}. 
Implementation details are presented in Web Appendix B, while details on block specification are described further in Section~\ref{sec:rda:implementation}.
These bootstrapping approaches generally maintain robustness and can be readily adapted for different modeling approaches and spatial structures, unlike commonly used alternatives like sandwich estimators (e.g., \cite{Hettinger2025MultiplyExposures}).

\subsection{Accounting for Other Drivers and Competition}\label{sec:methods:confound_comp}

When estimating the impact of a specific driver, we aim to isolate its effect from other factors including confounders (socioeconomic status, pre-tax prices, and pre-tax sales) and other drivers. 
Because drivers cannot be examined in isolation, we treat them as additional confounders based on the causal structures described in Figure~\ref{fig:conceptual}. 
In these cases, it is crucial to adjust for variables on alternative causal paths, including other drivers, even if they emerge after the intervention. 
However, adjusting for variables downstream of, or interacting with, the driver of interest can obscure its true effect by inadvertently controlling for part of its mechanism.

To assess cross-border shopping effects via border proximity, we estimate the $ADT$ while adjusting for pre-tax confounders ($\mathbf{X_i}$) and price competition from other Philadelphia stores, $h_{it}(\boldsymbol{\rho_t}|A=1)$.
However, we do not adjust for price changes, $P_i$, or price competition from nearby non-taxed stores, as cross-border shopping effects largely operate through these price disparities.

To evaluate the impact of economic competition, we estimate the $ADT$ while treating competition as the exposure of interest rather than a confounder. 
Here, we analyze a uni-dimensional driver ($\mathbf{D_{i}}=h_{i1}(\boldsymbol{\rho_1})$) while adjusting for confounders ($\mathbf{X_i}$).
To determine whether consumers respond to observed, rather than presumed, economic competition, we also adjust for border proximity ($B_i$) and store-level price changes ($P_i$). 

To evaluate how store-level price changes impact tax effectiveness, we estimate the $ADUTT$ while ensuring price changes are not conflated with pre-tax factors influencing store pricing. 
Therefore, we adjust for confounders ($\mathbf{X_i}$), border proximity ($B_i$), and pre-tax economic competition ($h_{i0}(\mathbf{\rho_0}))$. 
Here, we define our driver as a joint measure of price changes and nearby price competition ($\mathbf{D_{i}}=(P_i,h_{i1}(\mathbf{\rho_1}))$), to emulate the hypothetical scenario where stores randomly adjust prices but price changes maintain their correlation with price competition.

\section{Philadelphia Beverage Tax Analysis}\label{sec:rda}

\subsection{Data}

\begin{figure}[htbp]
\centering
\includegraphics[width=0.49\textwidth]{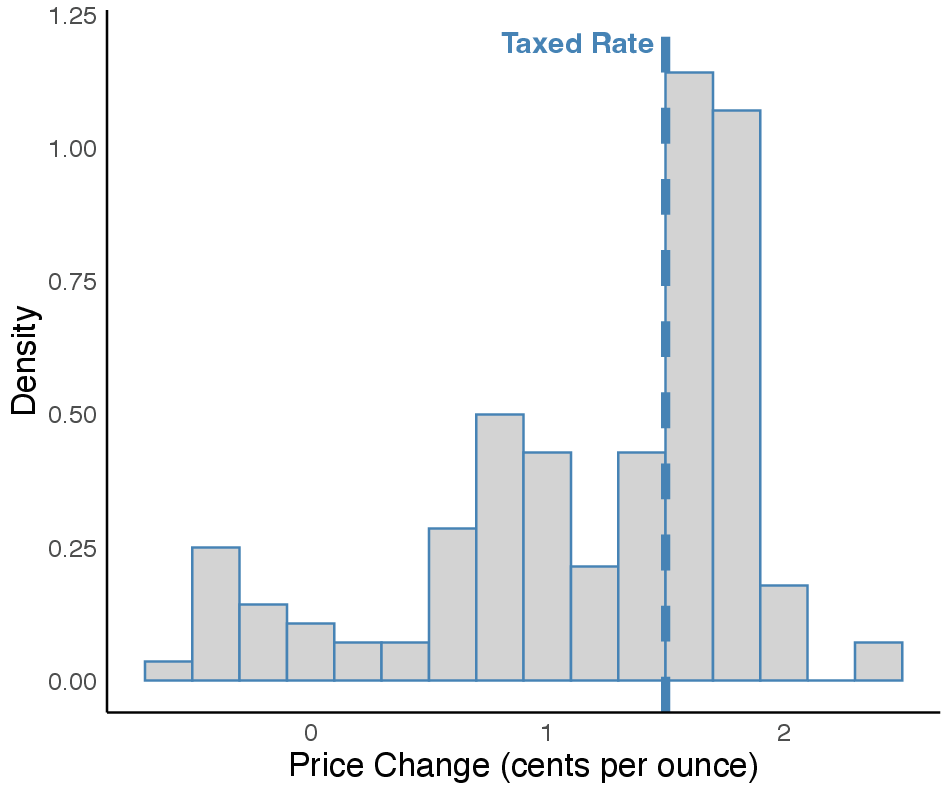}
\includegraphics[width=0.49\textwidth]{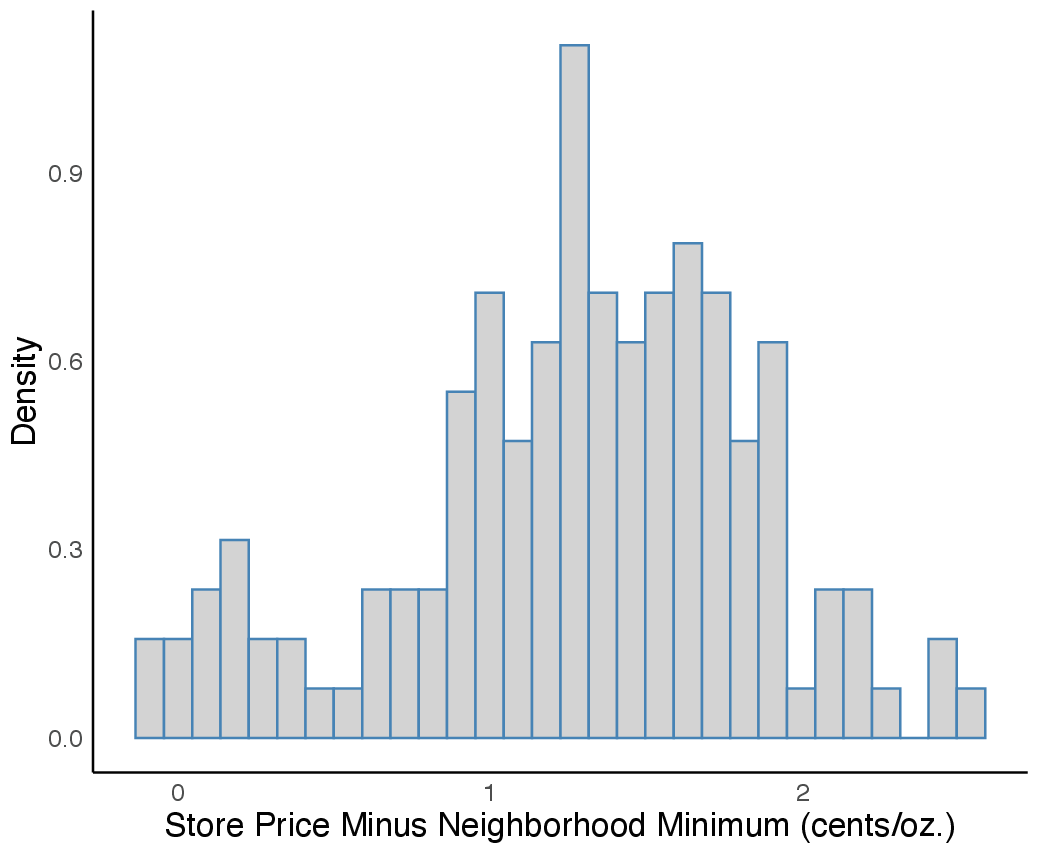}
\caption{Distribution of store-level price changes (left) and post-tax price competition (right) in Philadelphia.}\label{fig:price_change_hist}
\end{figure}

To evaluate drivers of effect heterogeneity in the Philadelphia beverage tax, we analyzed data from 140 Philadelphia pharmacies (treated group) and 123 pharmacies in Baltimore and non-neighboring Pennsylvania (PA) counties (control group). Additionally, we used price data from 32 pharmacies in neighboring non-taxed PA counties to measure economic competition. 
Data from Information Resources Inc. (IRI), previously described, included volume sales and taxed beverage prices aggregated over 4-week periods in the year before (2016) and after (2017) tax implementation ($m=1,...,13$)~\citep{MuthIRI2016, Roberto2019AssociationSetting}.

\begin{table}[htbp]
\caption{Mean (SD) of population characteristics for subgroups of Philadelphia stores based on border proximity (land distance to nearest non-taxed zip code) and control stores. Post-tax price competition among taxed stores ($h_{i1}(\boldsymbol{\rho_1} | A_i=1)$) does not exist for non-taxed stores.}\label{tab:dist_t1}%
\begin{tabular}{@{}l|ccc|c@{}}
\toprule
& \multicolumn{3}{c|}{Philadelphia} & \multicolumn{1}{c}{Control} \\
& $<3$ miles & $3-4.5$ miles & $>4.5$ miles & Baltimore \& Non-Border PA \\
Variable & $(n=42)$ & $(n=52)$ & $(n=46)$ & $(n=123)$ \\
\midrule
Pre-Tax Sales (x$10^3$oz.)        & 140.5 (83.7)  & 151.1 (86.3)  & 190.1 (100.2) & 99.5 (61.6)   \\
Pre-Tax Price (cents/oz.)        & 6.4 (0.6)     & 6.8 (0.7)     & 6.7 (0.7)        & 7.1 (0.7)     \\
Poverty Score      & 82.4 (11.0)     & 62.6 (24.8)   & 75.6 (22.5)   & 35.7 (32.9)   \\
Nonemployed Score      & 91.2 (14.4)   & 61.4 (27.0)       & 71.5 (28.3)       & 49.4 (28.5)   \\
$h_{i1}(\boldsymbol{\rho_1} | A_i=1)$ (cents/oz.)       & 1.05 (0.62)      & 1.24 (0.60)      & 1.19 (0.58)      & --   \\
\midrule
Change in Sales    & -40.2 (58.6)  & -29.7 (62.8)  & -35.2 (71.2)  & -10.8 (18.0)   \\
After Tax  (x$10^3$oz.) & & & & \\
\botrule
\end{tabular}
\end{table}

We then defined key measures as follows:
We calculated the land distance between the centroid of each Philadelphia zip code and the nearest non-taxed zip code as our measure of border proximity, $B_i$.
We calculated the average price of taxed beverages per unit per ounce at each store in each 4-week period as our measure of beverage price, $\rho_{itm}$. 
From this price measure, we calculated the difference between a store's average beverage price in the first three 4-week periods of 2017 and its 2016 average as our measure of price change, $P_{i}=\frac{1}{3}\sum\limits_{m=1}^{3} \rho_{i1m} - \frac{1}{13}\sum\limits_{m=1}^{13} \rho_{i0m}$.
Finally, we calculated the difference between the store's price and the minimum price at observed stores in the neighborhood of store $i$, $\mathcal{N}(i)$, as our measure of price competition, $h_{it}(\boldsymbol{\rho_{t}})=\frac{1}{3}\sum\limits_{m=1}^{3} \rho_{itm}-\min\limits_{j \in \mathcal{N}(i), j \neq i, m \in 1:3} \rho_{jtm}$.
Here, $\mathcal{N}(i)$ includes the store's zip code and adjacent zip codes.
For $h_{it}(\boldsymbol{\rho_{t}}|A=1)$, we used the same definition of $\mathcal{N}(i)$ but excluded neighboring non-taxed zip codes. 
We did not use prices from later periods of 2017 when defining $P_i$ and $h_{it}(\boldsymbol{\rho_t})$ as these may impose bias by representing post-tax sales adjustments. Distributions of store-level price changes and post-tax price competition in Philadelphia are depicted in Figure~\ref{fig:price_change_hist}.

We used average taxed beverage prices and sales per period in 2016, along with components of Social Deprivation Index (SDI) scores, as pre-tax confounders ($\mathbf{X_{im}}$) to account for population differences across stores exposed to different tax interventions or driver levels. 
SDI information derived from 2012-2016 American Community Survey (ACS) data was linked to stores at the zip-code level~\citep{TheRobertGrahamCenter2018SocialSDI}.
Pre-tax data revealed imbalances in confounder distributions across intervention status, border proximity, price changes, and economic competition (Tables~\ref{tab:dist_t1}-\ref{tab:price_t1}).

\subsection{Implementation}\label{sec:rda:implementation}

To implement our proposed estimators, we first modeled the four nuisance functions, $\mu_{0,\Delta}$, $\mu_{1,\Delta}$, $\pi_A$, and $\pi_D$. 
Since model misspecification is the primary source of bias when identification assumptions hold, we employed flexible machine learning estimators.
However, sample size constraints limited the use of certain non-Donsker class models which require sample-splitting to reduce bias due to overfitting~\citep{Naimi2023ChallengesAlgorithms,Balzer2023InvitedResearch}.

\begin{table}[htbp]
\caption{Mean (SD) of population characteristics for subgroups of Philadelphia stores based on store-level price changes in taxed beverages. At the 32 neighboring non-taxed PA stores, the average 2017 price was 7.0 cents per ounce (SD 0.7).}\label{tab:price_t1}%
\begin{tabular}{@{}l|ccccc@{}}
\toprule
& \multicolumn{5}{c}{Philadelphia Price Changes (cents/oz.)} \\
& $<0$ & $0-1$ & $1-1.5$ & $1.5-2$ & $>2$  \\
Variable & $(n=14)$ & $(n=32)$ & $(n=25)$ & $(n=66)$ & $(n=3)$ \\
\midrule
Pre-Tax Sales  (x$10^3$oz.)        & 228.8 (83.6)   & 193.1 (114.5)    & 162.5 (93.0)     & 130.3 (65.9)   & 153.1 (94.7)   \\
Pre-Tax Price (cents/oz.)        & 7.0 (0.4)        & 7.2 (0.5)      & 6.8 (0.8)      & 6.2 (0.4)      & 6.3 (0.5)      \\
Poverty Score       & 79.4 (19.2)    & 70.1 (20.8)      & 69.8 (23.1)    & 72.8 (23.0)      & 96.7 (0.5)     \\
Nonemployed Score       & 83.3 (23.1)    & 66.8 (29.3)      & 66.6 (28.8)    & 76.6 (25.4)      & 96.0 (2.9)     \\
$h_{i0}(\boldsymbol{\rho_0})$ (cents/oz.)        & 1.61 (0.37)     & 1.48 (0.42)      & 1.06 (0.65)      & 0.45 (0.36)      & 0.30 (0.4)      \\
\midrule
$h_{i1}(\boldsymbol{\rho_1})$ (cents/oz.)       & 0.25 (0.29)      & 1.30 (0.43)      & 1.50 (0.61)      & 1.37 (0.38)      & 1.94 (0.39)        \\
\midrule
Change in Sales      & 90.3 (80.8)    & -49.1 (67.0)   & -53.6 (48.5)   & -46.3 (26.2)   & -49.6 (28.9)   \\
After Tax (x$10^3$oz.) & & & & \\
\botrule
\end{tabular}
\end{table}

For $\mu_{0,\Delta}$, $\mu_{1,\Delta}$, and $\pi_A$, we used the SuperLearner algorithm with candidate learners including highly adaptive lasso (HAL), generalized additive models (GAM), and bayesian additive regression trees (BART)~\citep{vanderLaan2007SuperLearner, vanderLaan2017ALasso, Hastie2020Gam:Models, McCulloch2021BART:Trees}. 
While HAL and GAM belong to the Donsker class, BART does not; however, its probabilistic nature often mitigates overfitting in practice, making it highly effective for nuisance model estimation even without sample splitting~\citep{Dorie2019AutomatedCompetition}. 

Estimating the conditional density function, $\pi_D$, poses greater challenges as overly flexible models can introduce high variance and inflated weights.
To improve estimation efficiency, we used the HAL-based conditional density estimation procedure by  \cite{Hejazi2022Haldensify:R}.
For the joint price change and price competition exposure, we sequentially modeled $\pi_D$ due to the multi-dimensionality of $\mathbf{D_{i}}$. 
Specifically, we first modeled the density functions $p(P_i|\mathbf{X},A=1)$ and $p(h_{i1}(\boldsymbol{\rho_1})|P_i,\mathbf{X},A=1)$ separately using HAL, then multiplied them to obtain the estimated generalized propensity score, $p(P_i, h_{i1}(\boldsymbol{\rho_1})|\mathbf{X},A=1)$.
Each nuisance function model incorporated the pre-tax confounders and other driver measures described in Section~\ref{sec:methods:confound_comp} as $\mathbf{X}$.

To handle repeated sales observations, we matched each 2017 4-week period to the same period in 2016, creating a two-time period setting where time varied by tax period ($t$) but not seasonality ($m$). 
We then fit separate outcome models and propensity score models for each $m$ and estimated period-specific effects -- $ATT(m)$, $ADT(\boldsymbol{\delta},m)$, $ADUTT(m)$, and $REDA(m)$.
We averaged them over the last ten 4-week periods ($m=4,...,13$) as our total effect estimates to assess post-adjustment consumer behavior.
This approach accounts for seasonal sales patterns and avoids requiring parallel trends between consecutive 4-week periods~\citep{Hettinger2024DoublyInterference}.
Our estimates thus reflect average sales effects per store per 4-week period.

Beyond our proposed estimators, we computed two additional effect curves. 
First, when analyzing the cross-border shopping accessibility driver, we calculated placebo effect curves using pre-tax data ($m=4$ as control group, $m'=5,...,13$ as treated groups) as a limited proxy for counterfactual parallel trends. If these placebo estimates resemble actual effect curves, this may indicate confounding biases~\citep{Hettinger2025MultiplyExposures}.
Placebo tests for the price competition driver are not conducted as these are expected to yield non-null effects due to inherent associations between post-tax competition and pre-tax sales via their mutual associations with pre-tax competition.

Second, to illustrate potential differences with alternative approaches, we calculated \textbf{confounder-naive} effect curves, which estimate associations between post-tax price competition and post-tax sales without adjusting for other confounding factors. 
Specifically, we replaced $\widehat{\theta}_m(\mathbf{D})$ with a local linear kernel regression of $(Y_{1m}-Y_{0m})$ on $\mathbf{D}$ among treated units and estimated $\hat{\tau}$ as the sample mean of $Y_{1m}-Y_{0m}$ among control units, following \cite{Callaway2024Difference-in-differencesTreatment}.

For uncertainty quantification, we implemented our block bootstrapping approach, defining blocks similarly to our neighborhoods, $\mathcal{N}(i)$. 
Each of the $n_{zip}$ blocks corresponded to a Philadelphia zip code and its adjacent zip codes, thereby assuming that non-adjacent zip codes exert minimal influence on a store's prices and sales. 
The weighted block bootstrapping approach (Web Appendix B) accommodates these overlapping blocks.

\subsection{Results}

To align with our hypothetical experiments outlined in Section~\ref{sec:framing}, we estimated the $ATT$, which captures the overall tax effect independent of any drivers, along with the $ADT$ for border proximity and economic competition, and the $ADRTT$ and $REDA$ for price change.

The estimated ATT of the tax on Philadelphia pharmacies was -22.5 thousand ounces (95\% CI [-27.1, -18.0]) per store per 4-week period, indicating a substantial decline in sales due to the tax.

\begin{figure}[htbp]
\centering
\includegraphics[width=0.75\textwidth]{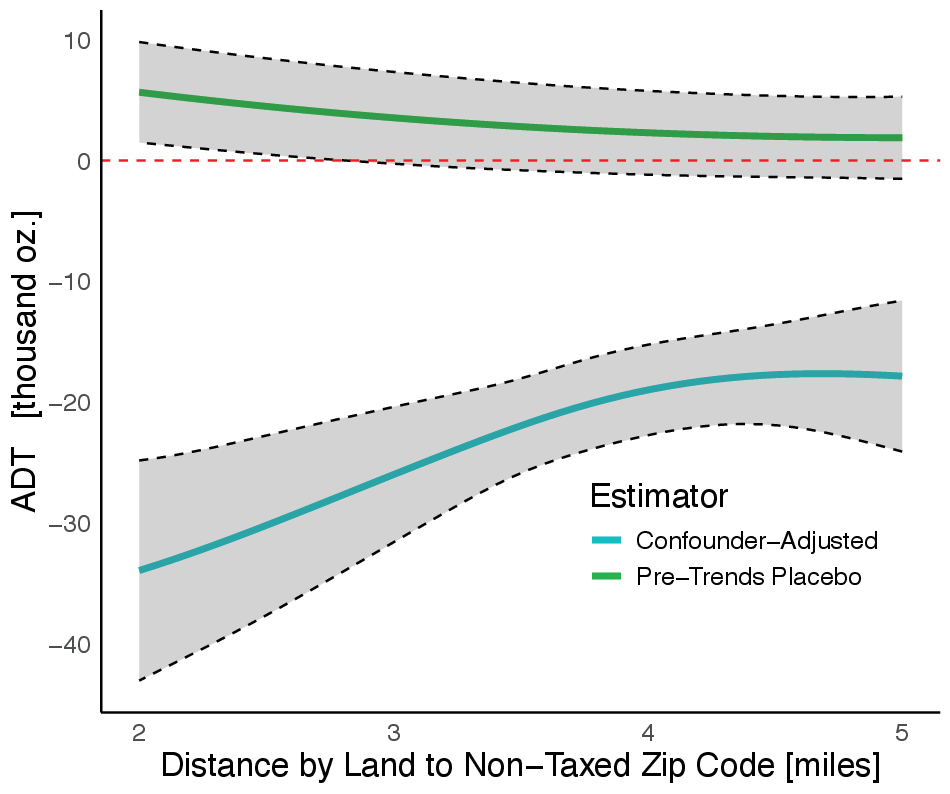}
\caption{Effect curves (ADT) for cross-border shopping accessibility along with pointwise 95\% confidence intervals. Curves represent the average effect of the Philadelphia beverage tax were all stores a certain distance to a non-taxed region. The top (green) curve represents the placebo pre-trends effect curve whereas the bottom (blue) curve represents the estimated ADT.}\label{fig:adt_dist}
\end{figure}

\subsubsection*{Cross-Border Shopping Accessibility}

A previous analysis of the ADT for border proximity examined its role in shaping tax effects but did not account for the influence of other drivers when isolating its effect~\citep{Hettinger2025MultiplyExposures}.

The ADT for border proximity (Figure~\ref{fig:adt_dist}) represents the average tax effect in Philadelphia if all stores were a fixed distance from a non-taxed store.
While the scenario where all stores had equal accessibility to a non-taxed store is hypothetical, this estimand also reflects the effect for stores $\delta$ miles from a non-taxed store, assuming they were representative (by observables) of the broader Philadelphia population.  

Tax effects varied significantly by distance, with cross-border accessibility influencing effectiveness up to approximately 4 miles, where the effect appeared to stabilize.
However, tax effects were significant across all distances.
Pre-trends placebo tests were flat and slightly above zero, indicating minimal bias and possible underestimation of the tax effect.

The estimated difference in sales between the $95^{\text{th}}$ percentile (5 miles) and $5^{\text{th}}$ percentile (2 miles) of distance was 16.1 thousand ounces (95\% CI [6.6, 25.5]), nearly half the effect size observed at 2 miles (-33.9 thousand ounces).
Notably, this was almost double the estimated difference using the confounding-naive approach defined in Section~\ref{sec:rda:implementation} (8.7 thousand ounces, 95\% CI [-1.4, 18.9]), suggesting confounders and other drivers may attenuate the observed relationship between distance and tax effects.

\begin{figure}[htbp]
\centering
\includegraphics[width=0.80\textwidth]{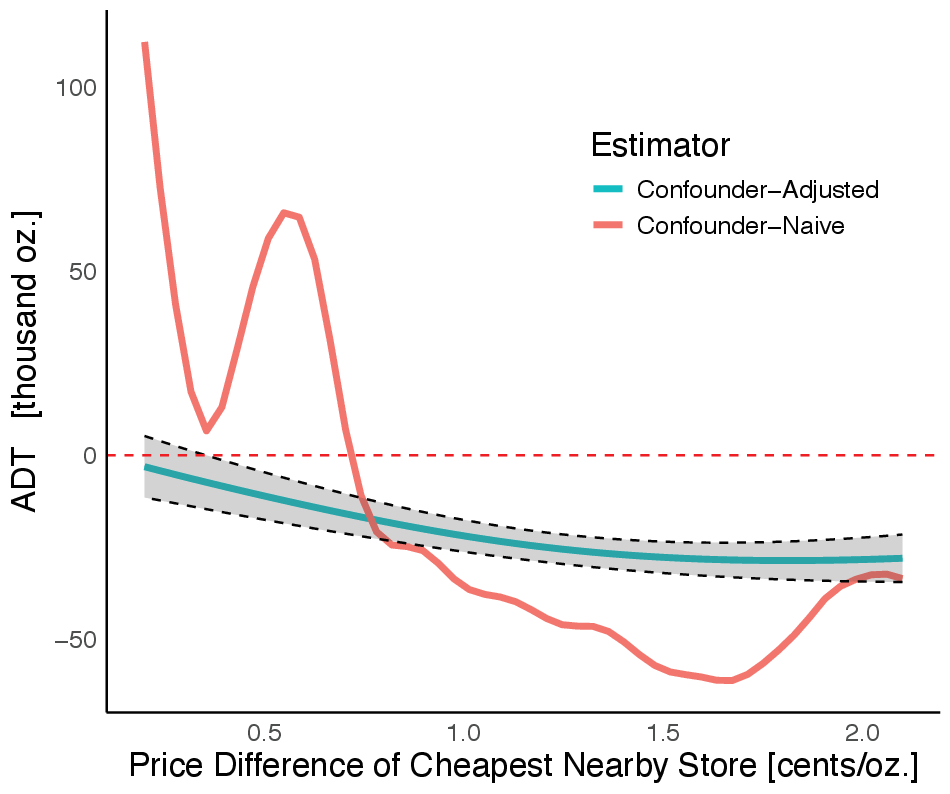}
\caption{Effect curves (ADT) for neighborhood price competition along with pointwise 95\% confidence intervals. Curves represent the average effect of the Philadelphia beverage tax were all stores in a neighborhood with a store that sold taxed beverages at a $\delta$ cent per ounce cheaper price. The more variable (red) curve represents the estimated effect curve without confounder adjustment, similar to the approach of \cite{Callaway2024Difference-in-differencesTreatment}, while the less variable (blue) curve represents the estimated effect curve after confounder adjustment, based on \cite{Hettinger2025MultiplyExposures}.}\label{fig:adt_comp}
\end{figure}

\subsubsection*{Economic Competition}

The ADT for economic competition (Figure~\ref{fig:adt_comp}) represents the tax effect in Philadelphia if all stores faced a fixed level of price competition from nearby stores.
Substantial heterogeneity suggests competition influenced tax effectiveness, with effects stabilizing near 1.5 cent-per-ounce price difference -- the tax rate itself.

Under low price competition, the estimated tax effect was near zero (-3.1 thousand ounces, 95\% CI  [-11.4, 5.2] at the $5^{\text{th}}$ percentile of competition), implying that the tax alone had little impact when post-tax prices remained similar across stores. 
However, this estimate extrapolates from limited data, as few stores exhibited both low competition and significant price changes.

Without adjusting for confounders, we would have estimated sales \textit{increases} under price differences less than 0.75 cents per ounce (Figure~\ref{fig:adt_comp}), a highly illogical conclusion. 
After adjusting for confounder, the estimated difference in sales between the $95^{\text{th}}$ percentile (2.1 cent per ounce difference) and $5^{\text{th}}$ percentile (0.2 cent per ounce difference) of competition was -24.9 thousand ounces (95\% CI [-35.8, -14.0]).

\subsubsection*{Price Change}

Table~\ref{tab:price_t1} shows a strong correlation between price change and post-tax price competition, with larger price increases linked to higher levels of price competition. This relationship further motivates our treatment of these factors as a joint driver.

The estimated ADUTT under an unconfounded price change was -25.0 thousand ounces (95\% CI [-28.7, -21.4]), suggesting a larger tax effect if store-level price changes were independent of observed covariates.

The REDA estimate was -11.1\% (95\% CI [-23.2\%, 1.1\%]), implying that the ATT would be 11.1\% larger if store price changes were not associated with socioeconomic factors and pre-tax sales, price, or price competition.
This suggests that how stores adjusted their prices may have dampened the tax's potential impact.

\section{Discussion}\label{sec:disc}

In this work, we leveraged recent DiD methodologies for continuous exposures to develop a robust framework for evaluating specific drivers of policy effect heterogeneity. 
A key contribution of our work is the conceptualization of relevant questions about policy effect heterogeneity through hypothetical experiments, where all units receive a certain driver level or driver levels are unconfounded.
This framework enables researchers to utilize causal estimands that distinguish heterogeneity due to observed confounders from heterogeneity driven by specific factors of interest.
Our estimation strategy extends existing methods to account for complex dynamics and adjust for confounding using flexible models, requiring only a subset of models to be correctly specified for consistency. 

Applying this framework to the Philadelphia beverage tax, we found evidence that cross-border shopping accessibility and price competition significantly shaped tax effects.
The impact of border proximity plateaued beyond four miles from a non-taxed store, while price competition effects stabilized a 1.5 cent-per-ounce price difference.
Additionally, we found some evidence that store-level pricing adjustments mitigated tax effects, possibly as a response to anticipated sales losses.
Our focus on Philadelphia pharmacies -- chosen for their large sample size and diverse exposure levels -- allowed for a detailed evaluation of these factors. 
However, consumer shifts from pharmacies to supermarkets due to higher pharmacy prices remain a plausible dynamic~\citep{Roberto2019AssociationSetting}.
Future work could explore this by incorporating supermarket prices into price competition measures or examining correlations between nearby supermarket and pharmacy sales.

While tailored to the Philadelphia beverage tax, our framework for defining key research questions within a causal framework and using rigorous methodologies to address complex policy dynamics has broader applications, including evaluations of vaccination, gun restriction, and abortion policies~\citep{Barkley2020CausalStudy, Raifman2020, Garnsey2021Cross-countryCare}.
Applying these methods in other contexts requires defining hypothetical experiments relevant to the policy question and identifying key dynamics to adjust for when isolating specific drivers.
Careful consideration of how interconnected factors evolve under these hypothetical settings is also essential. 
For example, when evaluating an ADT for changes in county vaccination rates (or beverage prices), researchers should not adjust for neighboring vaccination rates (or price competition) but rather set them realistically under the counterfactual scenario. 
This could be achieved by modeling both treated and neighboring exposures and evaluating them under alternative distributions reflecting the counterfactual scenario.

Despite its strengths, our approach relies on strong identification and modeling assumptions~\citep{Callaway2024Difference-in-differencesTreatment}. 
Continuous exposure measures require more stringent counterfactual parallel trends assumptions and more complex outcome and propensity score models than binary interventions. 
Sensitivity analyses are crucial to assess robustness to assumption violations and modeling choices, which remains an important area for future work~\citep{Hettinger2025MultiplyExposures}.

Future research could also incorporate network structures -- such as Philadelphia residents' travel patterns to non-taxed areas -- to refine effect estimates~\citep{Forastiere2024CausalInterference}.
However, these approaches often rely on strong assumptions are are complicated by incomplete store data. 
Addressing non-random missingness, when such information exists, could improve analyses even without incorporating full network structures.

Finally, our analysis focused on initial price changes, which may not capture long-term price fluctuations as stores adjust to shifts in demand. 
We used immediate post-tax price changes as an outcome unlikely to be influenced by post-tax sales, but future work could extend this using dynamic treatment regime frameworks to study evolving tax impacts on prices and consumer behavior~\citep{Chakraborty2014DynamicRegimes}.

\backmatter





\bmhead{Acknowledgements}

This work was supported by NSF Grant 2149716 (PIs: Mitra and Lee) and NIH Grant 1R01DK136515-01A1 (PI: Mitra).

\begin{appendices}


\label{sec:appendix}

\section*{Web Appendix A: Efficient Influence Functions}

The efficient influence function (EIF) for the $ATT$~\citep{Santanna2020} is given as:
\begin{equation*}
    \phi^{(ATT)}(\mathbf{X}, A, Y_0, Y_1) = \frac{A - \pi_A(\mathbf{X})}{P(A=1)(1-\pi_A(\mathbf{X}))}(Y_1 - Y_0 - \mu_{0, \Delta}(\mathbf{X}) )
\end{equation*}
The EIF for the $ADUTT$~\citep{Hettinger2025MultiplyExposures} is given as:
\begin{equation*}
    \phi^{(ADUTT)}(\mathbf{X}, A, \mathbf{D}, Y_0, Y_1) = \frac{A}{P(A=1)} \xi(\mathbf{X}, A, \mathbf{D}, Y_0, Y_1) -  \tau(\mathbf{X}, A, Y_0, Y_1) + J(\mathbf{X}, A)
\end{equation*}
Where 
\begin{align*}
    &\xi(\mathbf{X}, A, \mathbf{D}, Y_0, Y_1; \mu_{1,\Delta}, \pi_D) = m(\mathbf{D}|A=1) + \frac{(Y_1-Y_0) - \mu_{1,\Delta}(\mathbf{X}, \mathbf{D})}{\pi_D(\mathbf{X},\mathbf{D})} P(\mathbf{D}|A=1) \\
    &\tau(\mathbf{X}, A, Y_0, Y_1; \mu_{0,\Delta}, \pi_A) = \frac{A}{P(A=1)}\mu_{0,\Delta}(\mathbf{X}) + \frac{(1-A)\pi_A(\mathbf{X})[(Y_1 - Y_0) - \mu_{0,\Delta}(\mathbf{X})]}{P(A=1)(1-\pi_A(\mathbf{X}))} \\
    &J(\mathbf{X}, A; \mu_{1,\Delta}) = \frac{A}{P(A=1)} \int\limits_\mathbbm{D} \{ \mu_{1,\Delta}(\boldsymbol{\delta}, \mathbf{X}) - m(\boldsymbol{\delta}|A=1) \} dP(\boldsymbol{\delta}|A=1)\\
    &m(\mathbf{D}|A=1) = \int\limits_\mathbbm{X} \mu_{1,\Delta}(\mathbf{x}, \mathbf{D}) dP(\mathbf{x}|A=1)\\
    &P(\mathbf{D}|A=1) = \int\limits_\mathbbm{X} \pi_D(\mathbf{x},\mathbf{D}) dP(\mathbf{x}|A=1)
\end{align*}

Notably, the EIF for the ADT is not tractable as this functional is not pathwise differentiable without imposing parametric assumptions on the curve itself~\citep{Kennedy2024SemiparametricReview}. 
However, the EIF for the ADUTT can be used to robustly estimate the ADT~\citep{Hettinger2025MultiplyExposures}.
As alluded to in the main text, $J(\mathbf{X}, A)$ does not factor into point estimates for our estimands of interest.

\section*{Web Appendix B: Block Bootstrapping Approach}

Closed-form approaches are possible for $ATT$ and $ADUTT$, but do not carry the robustness properties of the estimators since they rely on parametric assumptions violated under misspecified models. 
Sandwich variance estimators are a happy medium between the computation time of bootstrap approaches and robustness limitations of standard closed form solutions, but require new algebraic definitions for each set of estimating models, which are not always possible. 
Instead, bootstrap approaches generally maintain robustness and can be adapted for different models as well as spatial structures, albeit under potentially computationally intensive procedures.
 with the bootstrapping approach.

Once blocks are defined, our procedure works as follows:
\begin{enumerate}
    \item Sample weights, $\gamma_b$, for each block $b=1,\dots,n_b$ from an independent distribution, i.e., $\gamma_b \sim \text{exponential}(1)$.
    \item For each store $i$, sum all of the weights pertaining to store $i$ as $\gamma_i = \sum\limits_{b=1}^{n_b} \mathbbm{1}\{i \in b\} \gamma_b$.
    \item Normalize weights so the average weight within each treatment group is one, i.e., $\gamma_i = (\gamma_i \sum\limits_{j=1}^n \mathbbm{1}\{A_j = A_i\}) / \sum\limits_{i=1}^n (\mathbbm{1}\{A_j = A_i\} \gamma_i)$.
    \item Plug weights into each step in the estimation process that relies on the sample.
    \begin{enumerate}
        \item[i.] In estimation step (1), fit models for $\hat{\mu}_{1,\Delta}$, $\hat{\mu}_{0,\Delta}$, $\hat{\pi}_A$,  and $\hat{\pi}_D$ with weights $\gamma_i$. 
        \item[ii.] In estimation step (2), when calculating $m(\mathbf{D}|A=1)$ and $P(\mathbf{D}|A=1)$, use a weighted empirical average.
        \item In estimation step (4), fit a weighted kernel regression with weights $\gamma_i$.
        \item In estimation step (5), use weighted empirical averages for empirical means.
    \end{enumerate}
    \item Repeat for each bootstrap sample and take the standard deviation ($\sigma$) of desired parameters for confidence intervals (i.e., add and subtract 1.96$\sigma$ from point estimate for 95\% CIs).
\end{enumerate}
The weighted block-sampling design comes with several benefits. 
First, it addresses potential spatial correlation within blocks. 
Second, store-level weights are constant over time, thereby addressing temporal correlation within stores.
Finally, by sampling continuous weights instead of discrete samples, this approach is more efficient for small samples by improving the observed support of $\mathbf{X}$ and $\mathbf{D}$ within given bootstrap samples. 

\end{appendices}


\bibliography{references}

\end{document}